\documentclass[
reprint,
bibnotes,
amsmath,amssymb,
aps,
prl,preprintnumbers
]{revtex4-2}

\usepackage{extarrows}
\usepackage{graphicx}
\usepackage{dcolumn}
\usepackage{enumitem}
\usepackage{bm}
\usepackage{float}
\usepackage[many]{tcolorbox}
\usepackage{shuffle}
\usepackage{environ}
\usepackage{physics}
\usepackage{tabularx}
\usepackage{diagbox}
\usepackage{subcaption}
\usepackage{multirow}

\usepackage{CJK}

\begin{document}

\begin{CJK*}{UTF8}{}
\CJKfamily{gbsn}
\preprint{MPP-2025-230}
\title{Positive Celestial Geometry}

\author{Jin Dong$^{1}$}
\email{jindong@mpp.mpg.de}
\author{Stephan Stieberger$^{1}$}
\email{stephan.stieberger@mpp.mpg.de \\}

\affiliation{$^{1}$Max-Planck-Institut f\"ur Physik, Werner-Heisenberg-Institut, Boltzmannstr. 8, 85748 Garching bei M\"unchen, Germany}

\begin{abstract}
Celestial amplitudes are multiple Mellin transforms w.r.t. conformal dimensions. For arbitrary multiplicity $n$ of massless states in sufficiently high space--time dimension $D$ we perform all Mellin integrations and find an associahedron description in celestial space. The latter expresses celestial tree--level $\phi^3$ amplitudes  as the canonical forms associated with this positive geometry. This yields a geometric interpretation of celestial amplitudes in terms of the underlying boundary geometry.
Our universal treatment of Mellin integrals in $D$ dimensions also provides a unified description of celestial amplitudes arising from different bulk theories.

\end{abstract}
\maketitle
\end{CJK*}

\section{Introduction}
Flat space--time holography aims to reformulate the bulk S-matrix of massless field theories in asymptotically flat space-time in terms of boundary theories living at null infinity. Two complementary frameworks for such a reformulation are Carrollian holography, based on the intrinsic Carrollian geometry of null--infinity  \cite{Barnich:2010eb,Bagchi:2010zz,Donnay:2022aba,Donnay:2022wvx} and celestial holography where scattering amplitudes are expressed as conformal correlators on the celestial sphere \cite{deBoer:2003vf,Strominger:2017zoo,Pasterski:2016qvg,Pasterski:2017kqt}.
In a mathematically rigorous top-down construction (as a special toy example, cf. e.g. \cite{Costello:2022jpg} for a self--dual theory on the four--dimensional Burns space) both the boundary theory and its correlators are specified purely from the geometry and kinematic structures of null infinity, independently of any initial presentation of the bulk S-matrix. Null infinity provides the natural arena for asymptotic in- and out-states of massless scattering, thereby linking bulk dynamics to boundary geometry.
Based on the ABHY associahedron of Arkani--Hamed, Bai, He and Yan~\cite{Arkani-Hamed:2017mur} in this work we develop a geometric construction of celestial amplitudes by realizing them as canonical differential forms on a dual kinematic space associated with null infinity. This dual space carries a natural positive--geometry structure, whose generalized associahedral facets encode the factorization and soft limits of the celestial amplitude. 

The Lorentz group ${\rm SO}(1, D\!-\!1)$ in $D$--dimensional Minkowski space acts as the Euklidean conformal group on coordinates  of the $(D\!-\!2)$--dimensional celestial sphere $S^{D-2}$.
Celestial amplitudes describe the on--shell asymptotic data of scattering processes in terms of states defined at past and future null infinity, without making any explicit reference to bulk space--time evolution. For massless $n$--point scattering in $D$ dimensional Minkowski space we write null momenta as
\begin{equation}\label{Momentum}
    p_i^\mu=\epsilon_i \,\omega_i\,q_i^\mu(\mathbf{z}_i)\;,\ i=1,\ldots,n\;,
\end{equation}
with energies $\omega_i$ and points $\mathbf{z}_i$ on the celestial sphere $S^{D-2}$ and $\epsilon_i=\pm1$ specifying in-- and out--states, respectively. For simplicity we take all  labels $\epsilon_i=+1$.
Amplitudes formulated w.r.t. the standard momentum eigenstate basis are converted into the boost eigenstate basis making conformal properties manifest. The $D$--dimensional celestial amplitude is  defined by Mellin transforming the momentum-space amplitude ${\cal A}_n (\omega_i, \mathbf{z}_i)$ over the energies $\omega_a$ with conformal weights $\Delta_a$
\begin{equation}\label{MellinT}
    \tilde{\cal A}_n(\Delta_i,\mathbf{z}_i) 
    = \prod_{a=1}^{n} \int_0^\infty \frac{d\omega_a}{\omega_a}\, \omega_a^{\Delta_a} \, \delta^{D}(\sum_{j=1}^n p_j) \,   {\cal A}_n (\omega_i, \mathbf{z}_i)\;,
\end{equation}
and transforms like a correlator of primary operators on a $(D\!-\!2)$-dimensional  conformal field theory (CFT). Generically, momentum conservation entering the Mellin integrals (\ref{MellinT}) implies delta-function support on certain patches of the celestial sphere \cite{Mizera:2022sln}. Performing explicitly the multiple Mellin integrals (\ref{MellinT}), which for $D\!=\!4$ represent some sort of  Aomoto--Gelfand integrals \cite{Schreiber:2017jsr,Hu:2021lrx},  is technically non--trivial. General expressions for $n\!=\!2,3$ and $n\!=\!4$ celestial scalar amplitudes in $D\geq 4$ dimensions have recently been considered in \cite{Kulkarni:2025qcx}. In the next section for general $n$ in $D\geq n-1$ we shall advance explicit analytic expressions for~(\ref{MellinT}). 

For massless external states, the relation between $D\!=\!4$ momentum space and celestial amplitudes is particularly simple.
The (complexified) points $z_i$ in the celestial sphere $S^{2}$ are related to the asymptotic directions of light-like momenta of external particles. The massless four-momentum (\ref{Momentum}) is parameterized by:
\begin{equation}\label{celcoord}
    q_i^\mu=  (1+z_i \bar{z}_i,\,z_i +\bar{z}_i\,,-i(z_i -\bar{z}_i)\,,1-z_i \bar{z}_i)\, ,
\end{equation}
with  $n$ points $z_i\in \mathbb{C}$ on the celestial sphere. We use $z_{i,j}:=z_i-z_j$ and $\bar z_{i,j}:=\bar z_i-\bar z_j$.

\section{Solving the Energy Scales and the Mellin integrals}

In this section, we show that in sufficiently high dimensions $D$ the product of energy scales takes a particularly simple form, which allows the Mellin integrations (\ref{MellinT}) to be evaluated explicitly. We first focus on the celestial ${\rm tr}(\phi^3)$ (or bi-adjoint $\phi^3$~\cite{Cachazo:2013hca,Cachazo:2013iea}) amplitude $\tilde{\cal A}^{\phi^3}_n(\Delta_i,\mathbf{z}_i)$ obtained from  the Mellin transform (\ref{MellinT}) of the momentum-space amplitude ${\cal A}^{\phi^3}_n (\omega_i, \mathbf{z}_i)$. However, we emphasize that the same method applies to general tree-level amplitudes, and the corresponding Mellin integrals can be evaluated in complete analogy. As we will explain later, the explicit realization of the underlying geometry is not possible in Lorentzian signature. We therefore work in the more general setting of ${\rm SO}(d, D-d)$ and in the Mellin transform~\eqref{MellinT} the celestial positions $\mathbf{z}_i$ collectively denote the variables $\vec{z}_i$ and $\vec{z}_i\!'$ in ${\rm SO}(d, D-d)$; in particular, for ${\rm SO}(1,3)$ these reduce to the usual celestial coordinates $z_i$ and $\bar{z}_i$. 

We now demonstrate how to obtain the solution for the energies $\omega$ and evaluate the Mellin integrals. The basic observation is that momentum conservation determines all energy ratios, leaving one overall scale unfixed. We write $\omega_a=\omega_n w_a(\mathbf{z}_i)$ for $a=1,\ldots,n-1$, with the integration domain giving the step functions $\Theta(w_a)$ below. The remaining components of momentum conservation impose a residual support on the celestial coordinates, denoted by $R_I(\mathbf{z}_i)=0$, and produce a local Jacobian ${\cal J}_I(\mathbf{z}_i)$; here $I$ labels the local choice of independent momentum-conservation components. Equivalently, the momentum-conservation delta function factorizes locally as
\begin{align}\label{eq: main delta factorized}
    \delta^D\!\left(\sum_{i=1}^n p_i\right)
    &=
    \frac{1}{\omega_n^{D-n+1}{\cal J}_I(\mathbf{z}_i)}
    \prod_{a=1}^{n-1}\delta\!\left(\omega_a-\omega_n w_a(\mathbf{z}_i)\right)
    \nonumber\\
    &\quad\times
    \delta^{(D-n+1)}\!\left(R_I(\mathbf{z}_i)\right)\, .
\end{align}
The explicit component solution, the Jacobian, and the residual delta functions are collected in the Supplemental Materials. We now provide the explicit solution for the energy scales. Let $Q$ be the $n\times n$ Gram matrix whose entries are given by inner products of the null direction vectors $q$, namely 
\begin{equation}
    Q_{i,j}\equiv q_i\cdot q_j\, .
\end{equation}
The relation $Q\omega=0$ is a direct consequence of momentum conservation: contracting $\sum_i\omega_i q_i^\mu=0$ with $q_j$ gives $\sum_i Q_{j,i}\omega_i=0$. The vector $\omega_i$ is therefore the null vector of $Q$. In a generic patch this null vector is one-dimensional, so $Q$ has rank $n-1$. The adjugate identity gives
\begin{equation}
    Q\,\operatorname{adj}(Q)=\det(Q)\,\mathbf{1}=0\,,
\end{equation}
then implies that every column of $\operatorname{adj}(Q)$ lies in the kernel of $Q$. Because this kernel is one-dimensional, the adjugate is a rank-one matrix built from the same null vector. Since $Q$ is symmetric, so is $\operatorname{adj}(Q)$, and therefore
\begin{equation} \label{eq: adj}
     \operatorname{adj}(Q)_{i,j} = \lambda  \omega_i \omega_j\,,
\end{equation}
for some scalar factor $\lambda$. Taking the only unfixed energy scale to be $\omega_n$, we have
\begin{equation} \label{eq: lambda}
\lambda=\frac{1}{\omega_n^2}\operatorname{adj}(Q)_{n,n}\equiv\frac{1}{\omega_n^2}\Lambda(\mathbf{z}_i)\,,
\end{equation}
where $\Lambda(\mathbf{z}_i):=\operatorname{adj}(Q)_{n,n}$. In this notation the energies are
\begin{equation} \label{eq: solution omega}
  \omega_a=\omega_n w_a(\mathbf{z}_i)=\omega_n\frac{\operatorname{adj}(Q)_{a,n}}{\Lambda (\mathbf{z}_i)}\,.
\end{equation}
Consequently, the Mandelstam variables take the form
\begin{equation} \label{eq: shat}
    s_{i,j}= 2p_i\cdot p_j =  \frac{\omega_n^2}{\Lambda(\mathbf{z}_i)}2\, Q_{i,j} \operatorname{adj} (Q)_{i,j} \equiv \frac{\omega_n^2}{\Lambda(\mathbf{z}_i)} \hat{s}_{i,j} \,,
\end{equation}
where we have defined
\begin{equation}\label{Qass}
\hat{s}_{i,j}:=2\, Q_{i,j}\; \operatorname{adj} (Q)_{i,j}\;.
\end{equation}
Note that the $\mathrm{tr}(\phi^3)$ amplitudes in momentum space are simple homogeneous rational functions of the Mandelstam variables. Therefore, our solution of the products of $\omega$'s trivializes the Mellin integrals, yielding
\begin{align}   \label{eq: celestial phi3}
    &\tilde{\cal A}^{\phi^3}_n(\Delta_i,\mathbf{z}_i) \nonumber \\
    =& \int\limits_0^\infty d\omega_n\, \frac{ \omega_n^{\sum_i \Delta_i-3(n-2)}}{\omega_n^{D-n+1} {\cal J}_I(\mathbf{z}_i)} \,
    \delta^{(D-n+1)}\!\left(R_I(\mathbf{z}_i)\right)
    \nonumber\\
    &\quad\times
    \prod_{a=1}^{n-1}\Theta(w_a) w_a^{\Delta_a-1}(\mathbf{z}_i)
    \Lambda^{n-3}(\mathbf{z}_i) \hat{{\cal A}}^{\phi^3}_n (\mathbf{z}_i) \nonumber \\
    =&  \frac{2\pi\delta\left(\sum\limits_{i=1}^n \Delta_i-2n-D+6\right)}
    {{\cal J}_I(\mathbf{z}_i)}
    \delta^{(D-n+1)}\!\left(R_I(\mathbf{z}_i)\right)\nonumber\\
    &\quad\times
    \prod_{a=1}^{n-1}\Theta(w_a) w_a^{\Delta_a-1}(\mathbf{z}_i)
    \Lambda^{n-3}(\mathbf{z}_i)\, \hat{{\cal A}}^{\phi^3}_n (\mathbf{z}_i)\,,
\end{align}
where ${\cal J}_I$ is the local Jacobian and $\delta^{(D-n+1)}(R_I(\mathbf{z}_i))$ imposes the residual support associated with the patch $I$. For $D=n-1$ this residual delta function is absent; for $D>n-1$ it gives the distributional support of the celestial amplitude on the celestial configuration space, see Eq.~\eqref{eq: celestial support}.

Strictly speaking the last equality with the delta function is obtained provided $\sum_{i=1}^n \Delta_i-2n-D+6\in i\mathbb{R}$. However, the underlying Mellin integral can be continued to non-zero real part \cite{ZagierMellin}. Quite nicely, $\hat{\mathcal{A}}^{\phi^3}_n(\mathbf{z}_i)$ is simply the original momentum-space amplitude with the replacement $s_{i,j} \to \hat{s}_{i,j}$. For instance:
\begin{equation*}
    \hat{{\cal A}}^{\phi^3}_4 (\mathbf{z}_i)= \frac{1}{\hat{s}_{1,2}} +\frac{1}{\hat{s}_{2,3}} \,,
\end{equation*}
\vspace{-0.5cm}
\begin{equation*}
\begin{aligned}
    \hat{{\cal A}}^{\phi^3}_5 (\mathbf{z}_i)
    =& \frac{1}{\hat{s}_{1,2}\hat{s}_{1,2,3}}+\frac{1}{\hat{s}_{2,3}\hat{s}_{2,3,4}}\\
    +&\frac{1}{\hat{s}_{3,4}\hat{s}_{3,4,5}}+\frac{1}{\hat{s}_{4,5}\hat{s}_{1,4,5}}+\frac{1}{\hat{s}_{1,5}\hat{s}_{1,2,5}} \,.
\end{aligned}
\end{equation*}
Let us illustrate the general construction at $n=4$ in ${\rm SO}(1,3)$. Here $D-n+1=1$, so after the three energy ratios are fixed one residual delta function remains. In the standard parametrization~\eqref{celcoord}, the energy-scale solution above gives
\begin{align}
    \Lambda(\mathbf{z}_i)
    &=16|z_{1,2}|^2|z_{1,3}|^2|z_{2,3}|^2\,,
    \nonumber\\
    w_1(\mathbf{z}_i)
    &=
    \frac{|z_{1,4}|^2|z_{2,3}|^2-|z_{1,3}|^2|z_{2,4}|^2-|z_{1,2}|^2|z_{3,4}|^2}
    {2|z_{1,2}|^2|z_{1,3}|^2}\,,
    \nonumber\\
    w_2(\mathbf{z}_i)
    &=
    \frac{|z_{1,3}|^2|z_{2,4}|^2-|z_{1,4}|^2|z_{2,3}|^2-|z_{1,2}|^2|z_{3,4}|^2}
    {2|z_{1,2}|^2|z_{2,3}|^2}\,,
    \nonumber\\
    w_3(\mathbf{z}_i)
    &=
    \frac{|z_{1,2}|^2|z_{3,4}|^2-|z_{1,4}|^2|z_{2,3}|^2-|z_{1,3}|^2|z_{2,4}|^2}
    {2|z_{1,3}|^2|z_{2,3}|^2}\, .
\end{align}
The derivation in the Supplemental Materials shows that the residual support is the familiar real-cross-ratio condition
\begin{equation}
    {\rm Im}\,r=0,\qquad
    r=\frac{z_{1,2}z_{3,4}}{z_{1,3}z_{2,4}}\, ,
\end{equation}
and that the local Jacobian combines with the residual delta function as in Eq.~\eqref{eq: four point delta support}. Substituting this result into the general formula~\eqref{eq: celestial phi3} gives
\begin{align}
    \tilde{\cal A}^{\phi^3}_4
    =&\,
    \frac{2\pi\delta\!\left(\sum_{i=1}^4\Delta_i-6\right)}
    {2|z_{1,3}z_{2,4}|^2}
    \delta\!\left({\rm Im}\,r\right)
    \nonumber\\
    &\quad\times
    \prod_{a=1}^{3}\Theta(w_a)w_a^{\Delta_a-1}
    \Lambda(\mathbf{z}_i)
    \left(\frac{1}{\hat{s}_{1,2}}+\frac{1}{\hat{s}_{2,3}}\right)\, .
\end{align}
This is the usual real-cross-ratio support of the four-dimensional celestial four-point amplitude, and agrees with the known results in the literature cf. e.g.~\cite{Pasterski:2017kqt,Nandan:2019jas,Mizera:2022sln} after matching conventions for the overall normalization.

In the following let us discuss the soft limit of our celestial scalar amplitude $\tilde{{\cal A}}^{\phi^3}_n$.
The leading single soft limit $p_n\rightarrow 0$ of the $\phi^3$ amplitude ${\cal A}^{\phi^3}_n$ in the momentum space basis ${\cal A}^{\phi^3}_n\rightarrow\left(\frac{1}{s_{1,n}}+\frac{1}{s_{n-1,n}}\right){\cal A}^{\phi^3}_{n-1}+\ldots$ translates into the celestial basis as 
\begin{equation}
 \lim_{\Delta_n\rightarrow1}   \tilde{{\cal A}}^{\phi^3}_n=
 \frac{\epsilon_n}{\Delta_n-1} \left(\frac{\epsilon_1\, e^{-\partial_{\Delta_1}}}{Q_{1,n}}+
 \frac{\epsilon_{n-1}\, e^{-\partial_{\Delta_{n-1}}}}{Q_{n-1,n}}\right)\tilde{{\cal A}}^{\phi^3}_{n-1} ,
\end{equation}
with operators $e^{-\partial_{\Delta}}$ 
shifting the dimension as $\Delta\!\to\!\Delta-1$ or equivalently, multiplying the celestial amplitude by $1/\omega$; here we have restored the in/out labels $\epsilon_i$. We emphasize that the soft limit can be derived in arbitrary dimensions. In particular, for $D\!=\!4$ we obtain the following expression in celestial coordinates~(\ref{celcoord}):
\begin{equation}
 \lim_{\Delta_n\rightarrow1}   \tilde{{\cal A}}^{\phi^3}_n=
 \frac{\epsilon_n}{1-\Delta_n} \left(\frac{\epsilon_1\, e^{-\partial_{\Delta_1}}}{2|z_{n,1}|^2}+
 \frac{\epsilon_{n-1}\, e^{-\partial_{\Delta_{n-1}}}}{2|z_{n,n-1}|^2}\right)\tilde{{\cal A}}^{\phi^3}_{n-1} .
\end{equation}
Note that the operator $e^{-\partial_{\Delta}}$ is related to the inverse action of the lightlike translation generator $P_+\!=\!\tfrac{1}{2}(P_0+P_3)$, which in the celestial representation is realized  by the shift operator $e^{\partial_\Delta}$ \cite{Stieberger:2018onx}. Thus, the soft scalar creates a non-holomorphic, weight-shifting insertion in the celestial correlator. This is to be contrasted with the leading soft limit of graviton and gluon amplitudes \cite{Strominger:2017zoo}.

\section{Celestial Associahedron}
In this section, we introduce the celestial associahedron. As we will see, by following exactly the same construction of the ABHY associahedron in kinematic space~\cite{Arkani-Hamed:2017mur}, the celestial associahedron yields a canonical form that corresponds to the colored amplitude of $\mathrm{tr}(\phi^3)$ in celestial coordinates.

The kinematic associahedron is constructed by $n(n-3)/2$ linearly independent planar variables defined as:
\begin{equation}
    X_{i,j}=(p_i+p_{i+1}+\ldots+p_{j-1})^2\,,
\end{equation}
with $X_{i,j}=X_{j,i}$ and $X_{i,i}=X_{i,i+1}=0$. The region in which all planar variables are positive is further constrained to an $(n-3)$-dimension subspace by imposing $(n-2)(n-3)/2$ chosen ABHY conditions:
\begin{equation}
    c_{i,j}=X_{i,j}+X_{i+1,j+1}-X_{i+1,j}-X_{i,j+1} >0 \,,
\end{equation}
where $c_{i,j}$ is the non-planar Mandelstam variables  $c_{i,j}=-s_{i,j}=-2p_i\cdot p_j$ for $i<j-1$. A convenient choice of coordinates for spanning the kinematic associahedron is $\{X_{1,3},X_{1,4},\ldots,X_{1,n-1}\}$, together with the ABHY conditions $\{c_{i,j}>0\}$ for $i,j \neq n$. Equivalently, one may choose $\{s_{1,2},s_{2,3},\ldots,s_{n-3,n-2}\}$ as an alternative set of coordinates under the same conditions. 

Before proceeding, we note that the construction above is unambiguous only when the spacetime dimension satisfies $D\geq n-1$, otherwise, the planar variables $X_{i,j}$ are subject to nontrivial Gram-determinant constraints. In addition, the imposed constraints require a specific sign pattern, namely $s_{i,i+1}>0$ and $s_{i,j}<0$ for $i<j-1$. It is straightforward to verify that such a sign pattern cannot be realized for $n\geq5$ in $\mathrm{SO}(1,D-1)$, even before specifying in/out states ({\it i.e.} the sign of $\epsilon_i$). The reason is simply that the $\mathrm{SO}(1,D-1)$ parametrization is equivalent to $q_i=(1,\hat{n}_i)$ with $\hat{n}$ to be a unit vector on $S^{D-1}$, which implies $q_i \cdot q_j=-1+\hat{n}_i \cdot \hat{n}_j  \leq 0$. Thus, in what follows we work in $\mathrm{SO}(d,D-d)$ with $d, D-d\geq 2$ and $ D\geq n-1$~\footnote{Although the geometry cannot be realized explicitly in Lorentzian signature, the resulting canonical form defined in other signatures can be analytically continued to the Lorentzian case.}.

\begin{figure*}
    \centering
    \includegraphics[width=0.32\textwidth]{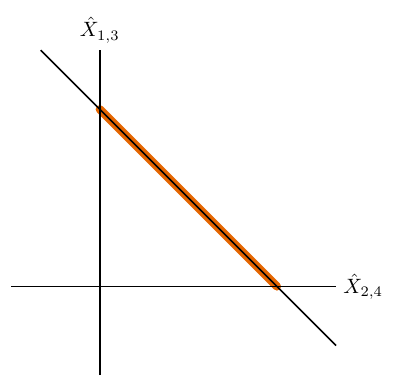}
    \includegraphics[width=0.32\textwidth]{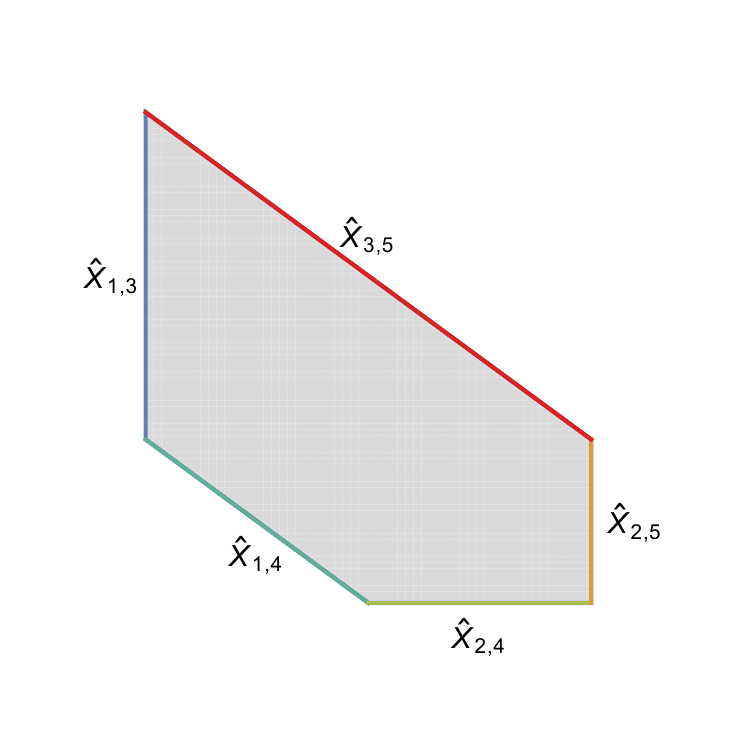}
    \includegraphics[width=0.32\textwidth]{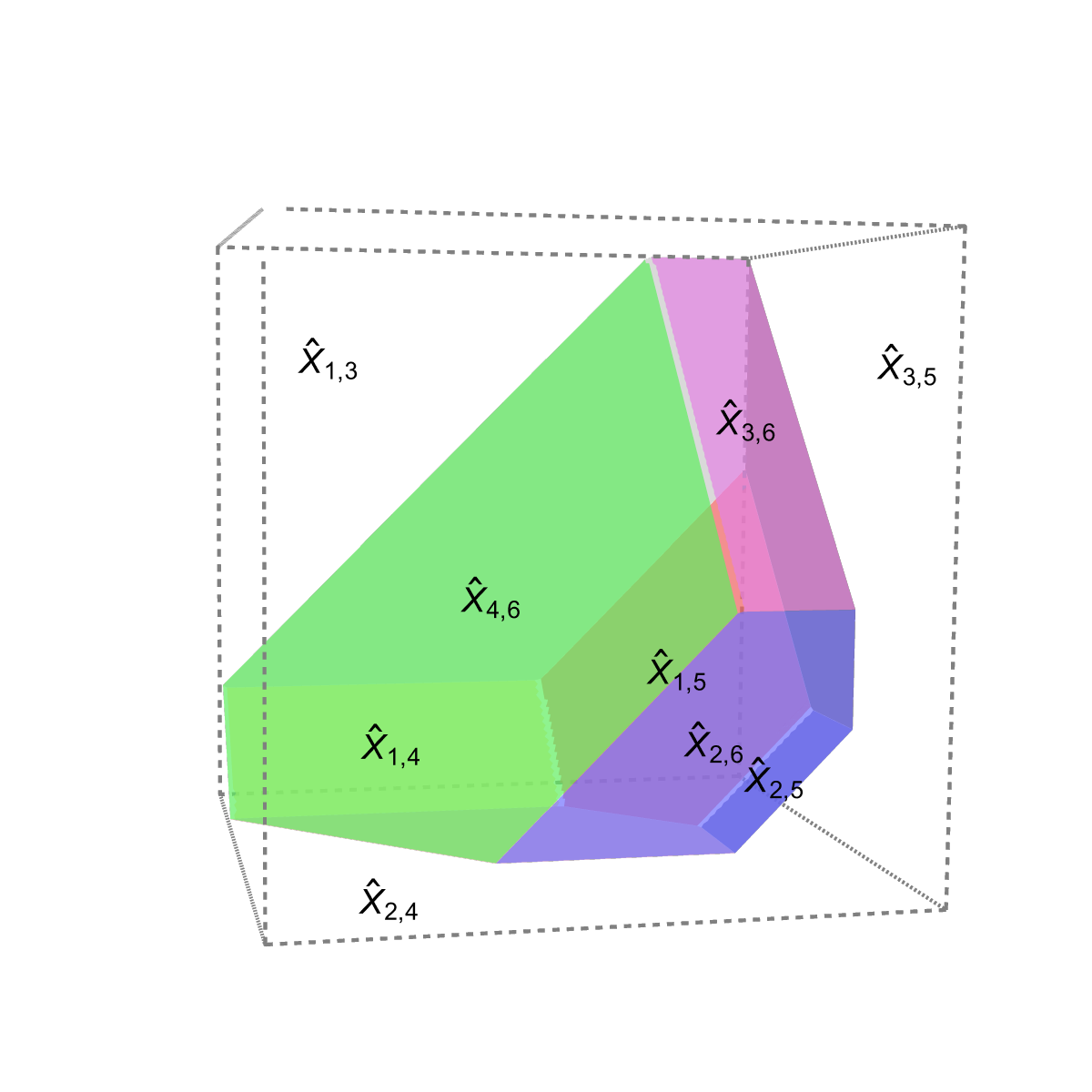}
    \caption{Celestial associahedra for $n=4,5$ and $n=6$.}
    \label{fig:Associahedra}
\end{figure*}

Since the original ABHY construction is formulated on the support of full momentum conservation, in what follows the celestial construction should be understood on the residual momentum-conservation support imposed by $\delta^{(D-n+1)}(R_I(\mathbf{z}_i))$. Given the results of the previous section, the most natural step is to define the celestial associahedron in complete analogy with the usual construction, simply replacing $s_{i,j} $ with its celestial counterparts $\hat{s}_{i,j}$ and define $\hat{c}_{i,j}=-\hat{s}_{i,j}$, $\hat{X}_{i,j}=\sum_{a\neq b,(a,b)\in \{i,i+1,\ldots,j-1\}} \hat{s}_{a,b}$. For example, at $n=4$ we have:
\begin{equation}
    \hat{X}_{1,3}=\hat{s}_{1,2}>0, \quad \hat{X}_{2,4}=\hat{c}_{1,3}-\hat{s}_{1,2}>0 \, \,,
\end{equation}
which corresponds to the line segment shown in Figure~\ref{fig:Associahedra}(left). For $n=5$ the celestial associahedron is described by the inequalities:
\begin{equation}
    \begin{cases}
        &\hat{X}_{1,3}= \hat{s}_{1,2}>0 \\
        &\hat{X}_{1,4}=-\hat{c}_{1,3}+\hat{s}_{1,2}+\hat{s}_{2,3}>0 \\
        &\hat{X}_{2,4}= \hat{s}_{2,3}>0 \\
        &\hat{X}_{2,5}= \hat{c}_{1,3}+\hat{c}_{1,4}-\hat{s}_{1,2}>0 \\
        &\hat{X}_{3,5}= \hat{c}_{1,3}+\hat{c}_{1,4}+\hat{c}_{2,4}-\hat{s}_{1,2}-\hat{s}_{2,3}>0\, , \\
    \end{cases}\quad
    \label{eq:5ptIneq}
\end{equation}
which is exactly the pentagon in middle of  Figure~\ref{fig:Associahedra}. A $3$-dimensional associahedron for $n=6$ is also displayed in Figure~\ref{fig:Associahedra} (right).

\noindent 

By construction, the canonical form of the celestial associahedron for $n=4$ reads~\cite{Arkani-Hamed:2017tmz,Arkani-Hamed:2017mur}:
\begin{equation}
    \hat{\Omega}_{4}^{\phi^3}=\frac{d \hat{s}_{1,2}}{\hat{s}_{1,2}}- \frac{d \hat{s}_{2,3}}{\hat{s}_{2,3}}=d \log (\frac{\hat{s}_{1,2}}{\hat{s}_{2,3}}) \,,
\end{equation}
In fact, on the celestial torus~\footnote{The parametrization we use for the celestial torus is presented in the Supplemental Materials; see also~\cite{Melton:2023bjw}.} with the conformal cross ratio $z=\frac{z_{1,2}z_{3,4}}{z_{2,3}z_{4,1}}=\frac{\hat{s}_{1,2}}{\hat{s}_{2,3}}\equiv\frac{\hat{X}_{1,3}}{\hat{X}_{2,4}}$ we have
\begin{equation}
     \hat{\Omega}_{4}^{\phi^3}=d \log z\ ,
\end{equation}
with the positivity constraint $z>0$. 

And for $n=5$, the canonical form reads:
\begin{align}
    \hat{\Omega}_{5}^{\phi^3}=& \frac{d\hat{s}_{1,2} \wedge d \hat{s}_{1,2,3}}{\hat{s}_{1,2}\hat{s}_{1,2,3}}+\frac{d\hat{s}_{2,3} \wedge d \hat{s}_{2,3,4}}{\hat{s}_{2,3}\hat{s}_{2,3,4}}+\frac{d\hat{s}_{3,4} \wedge d \hat{s}_{3,4,5}}{\hat{s}_{3,4}\hat{s}_{3,4,5}} \nonumber\\ 
    &+\frac{d\hat{s}_{4,5} \wedge d \hat{s}_{1,4,5}}{\hat{s}_{4,5}\hat{s}_{1,4,5}}+\frac{d\hat{s}_{1,5} \wedge d \hat{s}_{1,2,5}}{\hat{s}_{1,5}\hat{s}_{1,2,5}} \\
    =&d \log \frac{\hat{s}_{2,3}}{\hat{s}_{1,2}} \wedge d\log \frac{\hat{s}_{1,2}}{\hat{s}_{1,2,3}} +d \log \frac{\hat{s}_{1,2}}{\hat{s}_{2,3,4}} \wedge d\log \frac{\hat{s}_{2,3}}{\hat{s}_{3,4}}\,. \nonumber 
\end{align}
We then define the celestial scattering form as
\begin{equation} \label{eq: celestial form}
\begin{aligned}
    \tilde{\Omega}_{n}^{\phi^3}&=  \int_0^\infty d\omega_n\, \frac{\omega_n^{\sum_i \Delta_i-n}}{\omega_n^{D-n+1} {\cal J}_I(\mathbf{z}_i)} \,
    \delta^{(D-n+1)}\!\left(R_I(\mathbf{z}_i)\right)
    \\
    &\quad\times
    \prod_{a=1}^{n-1}\Theta(w_a) w_a^{\Delta_a-1}(\mathbf{z}_i)  \;\hat{\Omega}_{n}^{\phi^3} \\
    &= \frac{2\pi \delta(\sum_{i}\Delta_i-D)}{{\cal J}_I(\mathbf{z}_i)} \,
    \delta^{(D-n+1)}\!\left(R_I(\mathbf{z}_i)\right)
    \\
    &\quad\times
    \prod_{a=1}^{n-1}\Theta(w_a) w_a^{\Delta_a-1}(\mathbf{z}_i)  \;\hat{\Omega}_{n}^{\phi^3}\,. 
\end{aligned}
\end{equation}
We emphasize that $\tilde{\Omega}_{n}^{\phi^3}$ is the Mellin transform of the projective scattering form, not of the scalar amplitude as a function. Accordingly it is designed to encode the logarithmic singularities, residues, and factorization structure of the celestial associahedron. The scalar amplitude~\eqref{eq: celestial phi3} contains in addition the homogeneous factor $\Lambda^{n-3}(\mathbf{z}_i)/\omega_n^{2(n-3)}$, which is absent from the $d\log$ scattering form because the latter is projective, namely, the $d\log$ form is invariant under local ${\rm GL}(1)$ transformations $s_{i,j}\to\lambda' s_{i,j}$~\cite{Arkani-Hamed:2017mur}. It would be interesting to explore the implications of the resulting differences in the constraints on the conformal dimensions associated with~\eqref{eq: celestial phi3} and~\eqref{eq: celestial form}.

Let us also clarify the role of conformal covariance. Under a change of conformal section, $q_i\to\lambda_i q_i$, one has $Q_{i,j}\to\lambda_i\lambda_jQ_{i,j}$ and hence $\hat{s}_{i,j}=2Q_{i,j}\operatorname{adj}(Q)_{i,j}$ transforms by the same common factor $\prod_k\lambda_k^2$ for all pairs $(i,j)$. Thus ratios of the $\hat{s}_{i,j}$, and the $d\log$ canonical forms built from them, are conformally invariant. One may make this invariance manifest at the level of $\hat{s}_{i,j}$, for example by redefining $\hat{s}_{i,j}=\hat{s}_{i,j}/{\cal N}_{\rm cyc}$ with ${\cal N}_{\rm cyc}=\prod_a Q_{a,a+1}$, but this amounts to a choice of projective section and may introduce artificial singular loci. And as we explain below, the rescaling does not affect the geometry.

For $n=4$ and $n=5$ in the four--dimensional split signature $(--++)$, we also present the deformed geometry (parametrized by $\hat{s}_{ij}$ or equivalently $s_{ij}$) in terms of the celestial coordinates $\mathbf{z}_i$ as detailed in the Supplemental Materials. We refer the reader to Refs.~\cite{Atanasov:2021oyu,Casali:2022fro,Melton:2023bjw,Melton:2024jyq} for recent works on celestial amplitudes in split signature. Let us briefly discuss the dimension of the geometry in celestial coordinates, hereafter referred to as the deformed geometry. After quotienting out the redundancy associated with $\mathrm{SO}(d,D-d)$, the deformed geometry has dimension $n-3$:
\begin{equation}
\begin{aligned}
    &n(D-2)-(D-n+1)+1\\
    &\quad-\frac{(n-1)(2D-n)}{2}
    -\frac{(n-2)(n-3)}{2}=n-3\, .
\end{aligned}
\end{equation}
The above terms respectively account for the celestial coordinates, residual support constraints, overall scale, Lorentz-orbit dimension, and ABHY conditions. The Lorentz-orbit dimension takes into account the stabilizer acting on the orthogonal complement of the momentum span. We emphasize that the overall scale is essential for obtaining the correct dimension. It can be introduced through a common rescaling $s_{ij}=\rho\hat{s}_{ij}$, where $\rho$ is fixed by one of the ABHY conditions. Equivalently, the scale can be parametrized by the remaining energy variable $\omega_n$ through $\rho=\omega_n^2/\Lambda(\mathbf z_k)$.\footnote{To see that the choice of additional overall normalization is not important, let $\hat c_{ij}$ denote the ABHY combinations of $\hat X_{ij}$. Fixing $\rho$ through one condition, $\rho\hat c_{ab}=c_{ab}$, gives $\rho=c_{ab}/\hat c_{ab}$. Consequently, $s_{ij}=c_{ab}\hat s_{ij}/\hat c_{ab}$, which is invariant under any common nonvanishing rescaling $\hat s_{ij}\to f(\mathbf z_k)\hat s_{ij}$. The remaining ABHY conditions and positivity constraints are therefore independent of this normalization.}

We emphasize that our universal treatment of multi Mellin integrals can also be applied to a wide class of bulk theories, as illustrated by the (scalar-scaffolded) gluon and graviton amplitudes discussed in the Supplemental Materials.

\section{Celestial associahedron and string world--sheet}

The deformed celestial geometry is parameterized by the $\tfrac12 n(n-3)$  real positive parameters $\hat{X}_{ij}>0$ comprised by the celestial variables $\hat{s}_{ij}$ introduced in (\ref{Qass}).
It is straightforward to show that the latter
lead to the same set of  solutions $\sigma_l\in\mathbb{C}$ of the scattering equations \cite{Cachazo:2013gna} as the kinematic variables $s_{ij}$, i.e.
\begin{equation}\label{SQE}
\sum_{i\neq j}\frac{s_{ij}}{\sigma_i-\sigma_j}=0\ \  \Longleftrightarrow\sum_{i\neq j}\frac{\hat{s}_{ij}}{\sigma_i-\sigma_j}=0\ .
\end{equation}
Hence, according to \cite{Arkani-Hamed:2017mur} there is  a bijective map between
the $n\!-\!3$ positive celestial coordinates $\hat{X}_{ij}$ and a set of (real ordered) $n\!-\!3$ open string world--sheet coordinates $\{\sigma_l\}$. The latter represent one of the $(n-3)!$ saddle point solutions on the string world--sheet disk describing the high--energy limit of the tree--level open $n$--point string amplitude. This way for any $n$ the set of equations (\ref{SQE}) yields a bijective map from the positive geometry of celestial associahedron (described by the positive coordinates $\hat{X}_{ij}$) to the geometry of open string world--sheet (described by the Riemann sphere of $n\!-\!3$ marked points $\sigma_l$), i.e. points of the celestial associahedron are mapped to vertex operator positions $\{\sigma_l\}$ along the boundary of the disk. Therefore, (\ref{SQE}) provides  a string interpretation of our celestial associahedron mapping the interior of the celestial associahedron to the  world--sheet associahedron. 

On the other hand, in \cite{Stieberger:2018edy,Kervyn:2025wsb,Castiblanco:2024hnq} it has been demonstrated that for $n\!=\!4$ and $n\!=\!5$ the infinite energy limit of the celestial string amplitude corresponds to the high--energy limit of that  amplitude.
This gives a direct connection between the celestial sphere and the string world--sheet relating points on the celestial sphere to saddle points on the string world--sheet. Since the latter are determined~\footnote{Note, that for $n=5$ there are two sets of solutions of (\ref{SQE}), which on the celestial sphere are related through the anti--podal map.} by (\ref{SQE}) we believe that the celestial associahedron allows to generalize this relation to generic $n$. At any rate this construction offers a concrete model illustrating how boundary conformal data at null infinity may arise from a fundamentally geometric and worldsheet-like description, strengthening the top-down perspective on celestial holography.


\section{Concluding remarks}
We have shown that for arbitrary numbers $n$, celestial $n$--point $\phi^3$ amplitudes in $D$ dimensions admit a purely geometric formulation: they arise as canonical differential forms on a dual kinematic space naturally associated with null infinity. This space carries a positive-geometry structure whose generalized associahedral facets encode the factorization channels and soft limits of massless scattering. In the celestial associahedron, the positivity of the dual kinematic space and the choice of associahedral boundaries select the admissible celestial configurations, thereby geometrizing the massless scattering. The resulting celestial associahedron furnishes a top-down geometric encoding of the kinematic and factorization data at null infinity. Its canonical form $\hat{\Omega}_n$ organizes the singularity and factorization structure of celestial amplitudes. In this sense, the celestial associahedron constitutes a key structural ingredient in a prospective top--down formulation of celestial holography, in which boundary geometry and positive geometry organize the holographic data.

More broadly, our universal treatment of multi Mellin integrals and the positive celestial geometry give a connection to the (open) string world--sheet and suggest a unified framework for celestial amplitudes for a wide class of bulk theories.

A first open question is to clarify the relation between the celestial associahedron and the data of a putative celestial conformal field theory (CCFT)/Carrollian theory at null infinity. Recent work has made the relation between Carrollian amplitudes, celestial symmetries, and BMS fluxes increasingly explicit: Carrollian amplitudes provide position-space correlators at ${\cal I}$, with celestial symmetry algebras acting on Carrollian operators, while BMS fluxes are related to celestial currents such as the supertranslation operator and the celestial stress tensor \cite{Donnay:2021wrk,Mason:2023mti}. It would therefore be natural to ask whether the BMS/Carrollian Ward identities can be represented as differential constraints on the canonical form of the celestial associahedron, with facets or boundary degenerations corresponding to soft insertions, charge fluxes, and factorization channels. Such a correspondence would tie the positive-geometry picture more tightly to the symmetry and operator content of celestial holography. In this sense, the celestial associahedron should not merely be viewed as a positive geometry for scalar celestial amplitudes, but as a possible geometric arena for the Ward identities and asymptotic charges of flat-space holography. Establishing such a dictionary would provide a direct bridge between positive geometry and the symmetry algebra of null infinity.

A second direction concerns the celestial operator product expansions (OPEs) which are controlled by the collinear limits of the underlying celestial amplitudes and thus by the factorization properties of the celestial associahedron. In particular, the leading OPE singularities are in one--to--one correspondence with specific facets of the positive geometry, so that OPE coefficients are directly encoded in residues of the canonical form. 
This provides a geometric organizing principle for celestial operator products and suggests that more general CCFT data --such as higher--point OPEs and conformal blocks-- should likewise admit an interpretation in terms of degenerations and intersections of celestial positive geometries.

\section{Acknowledgments}
It is our pleasure to thank Nima Arkani-Hamed, Qu Cao, Carolina Figueiredo, Song He and Tom Taylor for discussions. 
This work is supported by the DFG grant 508889767 {\it 
Forschungsgruppe ``Modern foundations of scattering amplitudes''}.

\newpage

\onecolumngrid
\appendix
\setcounter{equation}{0}
\renewcommand{\theequation}{S\arabic{equation}}

\vskip1cm
\begin{center}
    \textbf{\Large Supplemental Materials}
\end{center}

\section{Jacobian and Residual Momentum-Conservation Support}\label{app:jacobian-support}
In this section of the Supplemental Materials we give the details of the local momentum-conservation analysis used in the main text. Let
\begin{equation}
    \sum_{i=1}^n \omega_i V_i^\mu=0\,,
    \qquad V_i^\mu=q_i^\mu(\mathbf{z}_i)\, .
\end{equation}
At a generic point the rank is $n-1$, so the energy ratios are fixed while one overall scale remains. We choose $\omega_n$ as this scale and choose a local patch $I$ consisting of $n-1$ independent component equations with spacetime indices $\mu_1,\ldots,\mu_{n-1}$. With
\begin{equation}
    B_I(\mathbf{z}_i)_{\alpha a}=V_a^{\mu_\alpha}\,,
    \qquad a,\alpha=1,\ldots,n-1\,,
\end{equation}
the chosen components of momentum conservation are
\begin{equation}
    \sum_{a=1}^{n-1}B_I(\mathbf{z}_i)_{\alpha a}\omega_a
    =
    -\omega_n V_n^{\mu_\alpha}\, .
\end{equation}
Therefore
\begin{align}
    \omega_a&=\omega_n w_a(\mathbf{z}_i)\,,\nonumber\\
    w_a(\mathbf{z}_i)&=-\sum_{\alpha=1}^{n-1}
    (B_I^{-1})_{a\alpha}V_n^{\mu_\alpha}\,,
    \qquad a=1,\ldots,n-1\, .
\end{align}
The Jacobian for these $n-1$ energy delta functions is
\begin{equation}
    {\cal J}_I(\mathbf{z}_i)=\left|\det B_I(\mathbf{z}_i)\right|\, .
\end{equation}

The remaining $D-n+1$ components do not determine any further energy scale. Instead, after substituting the solution for the energy ratios, they impose the residual support constraints
\begin{equation}\label{eq: residual constraints}
    R_I^\rho(\mathbf{z}_i):=
    V_n^\rho+\sum_{a=1}^{n-1}w_a(\mathbf{z}_i)V_a^\rho=0\,,
    \qquad \rho\notin I\, .
\end{equation}
Here $\rho\notin I$ denotes a spacetime component not among the chosen components $\mu_1,\ldots,\mu_{n-1}$. Equivalently, these constraints are the vanishing of the $n\times n$ minors obtained by adjoining one remaining row $\rho$ to the chosen set $I$:
\begin{equation}\label{eq: residual minors}
    R_I^\rho(\mathbf{z}_i)
    =
    \frac{\det V_{I\cup \{\rho\}}}{\det B_I}\,,
    \qquad \rho\notin I\, .
\end{equation}
Indeed, writing the $n\times n$ matrix with rows $I\cup\{\rho\}$ in block form gives
\begin{equation}
    V_{I\cup\{\rho\}}
    =
    \begin{pmatrix}
        B_I & V_n^I\\
        V^\rho & V_n^\rho
    \end{pmatrix},
\end{equation}
where $(V_n^I)_\alpha=V_n^{\mu_\alpha}$ and $(V^\rho)_a=V_a^\rho$. The Schur complement gives
\begin{equation}
    \det V_{I\cup\{\rho\}}
    =
    \det B_I\,
    \left(V_n^\rho-V^\rho B_I^{-1}V_n^I\right)
    =
    \det B_I\,R_I^\rho(\mathbf{z}_i)\,,
\end{equation}
which proves Eq.~\eqref{eq: residual minors}.

With $\omega_n>0$, the momentum-conservation delta function factorizes locally as
\begin{align}\label{eq: delta factorized}
    \delta^D\!\left(\sum_{i=1}^n \omega_i V_i\right)
    &=
    \frac{1}{\omega_n^{D-n+1}{\cal J}_I(\mathbf{z}_i)}
    \prod_{a=1}^{n-1}\delta\!\left(\omega_a-\omega_n w_a(\mathbf{z}_i)\right)
    \nonumber\\
    &\quad\times
    \prod_{\rho\notin I}\delta\!\left(R_I^\rho(\mathbf{z}_i)\right)\, .
\end{align}
The factor $\omega_n^{-(D-n+1)}$ comes from the remaining components, since after the first $n-1$ delta functions are used each remaining equation becomes $\omega_n R_I^\rho(\mathbf{z}_i)=0$. The integration range $\omega_a>0$ also gives the positivity conditions
\begin{equation}\label{eq: celestial support}
    R_I^\rho(\mathbf{z}_i)=0\,,
    \qquad \rho\notin I\,,
    \qquad
    w_a(\mathbf{z}_i)>0\,,
    \qquad a=1,\ldots,n-1\, .
\end{equation}
The residual delta function used in the main text is therefore
\begin{equation}
    \delta^{(D-n+1)}\!\left(R_I(\mathbf{z}_i)\right):=
    \prod_{\rho\notin I}\delta\!\left(R_I^\rho(\mathbf{z}_i)\right)\, .
\end{equation}

One may also solve the residual delta functions locally in celestial coordinates. We write a local split schematically as $\mathbf{z}_i=(x,u)$, where the variables $x$ and $u$ are local functions of the celestial coordinates $(\vec{z}_i,\vec{z}_i\!')$ in the general ${\rm SO}(d,D-d)$ parametrization, or of $(z_i,\bar z_i)$ for ${\rm SO}(1,3)$. If $u^r$, $r=1,\ldots,D-n+1$, are chosen transverse to the support with
\begin{equation}
    \det\!\left(\frac{\partial R_I^{\rho_r}}{\partial u^s}\right)\neq0\,,
\end{equation}
then the implicit-function theorem gives local branches $u=u_\ell(x)$ satisfying $R_I^\rho(x,u_\ell(x))=0$. Hence
\begin{equation}\label{eq: solve residual delta}
    \delta^{(D-n+1)}\!\left(R_I(\mathbf{z}_i)\right)
    =
    \sum_\ell
    \frac{\prod_{r=1}^{D-n+1}\delta\!\left(u^r-u^r_\ell(x)\right)}
    {\left|\det\left(\partial R_I^{\rho_r}/\partial u^s\right)\right|_{u=u_\ell(x)}}\, .
\end{equation}
This is the most explicit form of the residual delta functions that does not require choosing a special signature or a special parametrization. To write the branches $u_\ell(x)$ themselves in closed form one must specify the corresponding celestial coordinates, since the choice of transverse variables depends on the coordinate patch.

Let us spell this out for the four-point example in ${\rm SO}(1,3)$. In the standard parametrization~\eqref{celcoord}, we make the constant change of momentum components
\begin{equation}
    \begin{pmatrix}
    \tilde q_i^0\\
    \tilde q_i^1\\
    \tilde q_i^2\\
    \tilde q_i^3
    \end{pmatrix}
    =
    \begin{pmatrix}
    (q_i^0+q_i^3)/2\\
    (q_i^1+iq_i^2)/2\\
    (q_i^1-iq_i^2)/2\\
    (q_i^0-q_i^3)/2
    \end{pmatrix}
    =
    \begin{pmatrix}
    1\\
    z_i\\
    \bar z_i\\
    z_i\bar z_i
    \end{pmatrix}.
\end{equation}
The Jacobian of this change of components is a constant independent of $\mathbf{z}_i$, and hence can be absorbed into the overall normalization. Momentum conservation is therefore equivalently written as $\sum_i\omega_i(1,z_i,\bar z_i,z_i\bar z_i)=0$. Choosing the first three components defines the patch $I_4$, with
\begin{equation}
    B_{I_4}=
    \begin{pmatrix}
        1&1&1\\
        z_1&z_2&z_3\\
        \bar z_1&\bar z_2&\bar z_3
    \end{pmatrix},
    \qquad
    {\cal J}_{I_4}=
    \left|z_{1,2}\bar z_{1,3}-z_{1,3}\bar z_{1,2}\right|\, .
\end{equation}
The remaining component gives
\begin{equation}\label{eq: four point residual}
    R_{I_4}(\mathbf{z}_i)
    =
    \frac{z_{1,2}z_{3,4}\bar z_{1,3}\bar z_{2,4}
    -\bar z_{1,2}\bar z_{3,4}z_{1,3}z_{2,4}}
    {z_{1,2}\bar z_{1,3}-z_{1,3}\bar z_{1,2}}\, .
\end{equation}
Defining
\begin{equation}
    r=\frac{z_{1,2}z_{3,4}}{z_{1,3}z_{2,4}}\,,
\end{equation}
the numerator of Eq.~\eqref{eq: four point residual} is
\begin{equation}
    z_{1,2}z_{3,4}\bar z_{1,3}\bar z_{2,4}
    -\bar z_{1,2}\bar z_{3,4}z_{1,3}z_{2,4}
    =
    2i\,|z_{1,3}z_{2,4}|^2\,{\rm Im}\,r\, .
\end{equation}
Since the denominator has absolute value ${\cal J}_{I_4}$, the product of the local Jacobian and the residual delta function becomes
\begin{equation}\label{eq: four point delta support}
    \frac{1}{{\cal J}_{I_4}(\mathbf{z}_i)}\,\delta(R_{I_4})
    =
    \frac{1}{2|z_{1,3}z_{2,4}|^2}\,
    \delta\!\left({\rm Im}\,r\right)\, ,
\end{equation}
up to the same constant normalization from the initial change of momentum components. Thus the residual momentum-conservation support is the usual real-cross-ratio condition ${\rm Im}\,r=0$.

\section{Deformed geometry in the celestial torus}
In this section, we present the deformed geometry expressed in celestial coordinates. As noted earlier, the construction is well-defined in $\mathrm{SO}(d,D-d)$ with $D\geq n-1$ and $d, D-d\geq 2$. Therefore we study the 4- and 5-point geometries at four--dimensional split signature $(--++)$. These examples should be viewed as concrete split-signature illustrations of the general construction on the residual momentum-conservation support, rather than as an additional derivation of the support constraints. We parametrize the null direction vector on the celestial torus as
\begin{equation}
    q_i^\mu=  (1-z_i z_i',\,z_i+z_i'\,,1+z_i z_i'\,,z_i-z_i')\, ,
\end{equation}
where $z_i$ and $z_i'$ are independent real variables. This gives
\begin{equation}
    q_i \cdot q_j =2 z_{i,j} z_{i,j}'
\end{equation}
where $z_{i,j}':=z_i'-z_j'$. Meanwhile, the $\mathrm{SO}(2,2)$ redundancy is decomposed into $\mathrm{SL}(2,\mathbb{R})\times \mathrm{SL}(2,\mathbb{R})$. Accordingly, we fix 3 of the $z$ and $z'$ coordinates by choosing
\begin{equation}
    (z_1,z_{n-1},z_{n})=(z_1',z_{n-1}',z_{n}')=(0,1,2) \,.
\end{equation}
As discussed in the main text, the celestial associahedron is equivalent to the kinematic associahedron. In this section of the Supplemental Material, we derive the deformed geometry by solving the constraints on celestial coordinates read from the kinematic variables $s_{i,j}$.

For $n=4$, four--dimensional momentum conservation allows us to solve for four variables, for examples, $\omega_1,\omega_2,\omega_3$ and $z_2'$. We then fix $\omega_4$ using the ABHY condition:
\begin{equation}
    s_{1,3}= \frac{\omega_4^2}{\operatorname{adj} (Q)_{4,4}}2\, Q_{1,3} \operatorname{adj} (Q)_{1,3} = -c_{1,3} \, ,
\end{equation}
where $c_{1,3}$ is taken to be a positive constant. In the end, the dependence of $s_{1,2}$ on $z_2$ is quite simple:
\begin{equation}
    s_{1,2}=\frac{c_{1,3} z_2}{2-z_2}, 
\end{equation}
Thus, the original line segment
\begin{equation}
    0<s_{1,2}<c_{1,3} 
\end{equation}
is mapped to
\begin{equation}
   0<z_2<1\,,
\end{equation}
which remains a line segment. The canonical form associated with this line segment reads:
\begin{equation}
    \Omega_4= d \log(\frac{z_2}{1-z_2})\, ,
\end{equation}
With the same gauge choice, this expression is precisely the celestial scattering form
\begin{equation}
\hat{\Omega}_{4}^{\phi^3}
=d\log z=d\log\left(\frac{z_{1,2}z_{3,4}}{z_{2,3}z_{4,1}}\right)
\end{equation}
up to an overall normalization.

\begin{figure*}[h]
    \centering
    \includegraphics[width=0.33\textwidth]{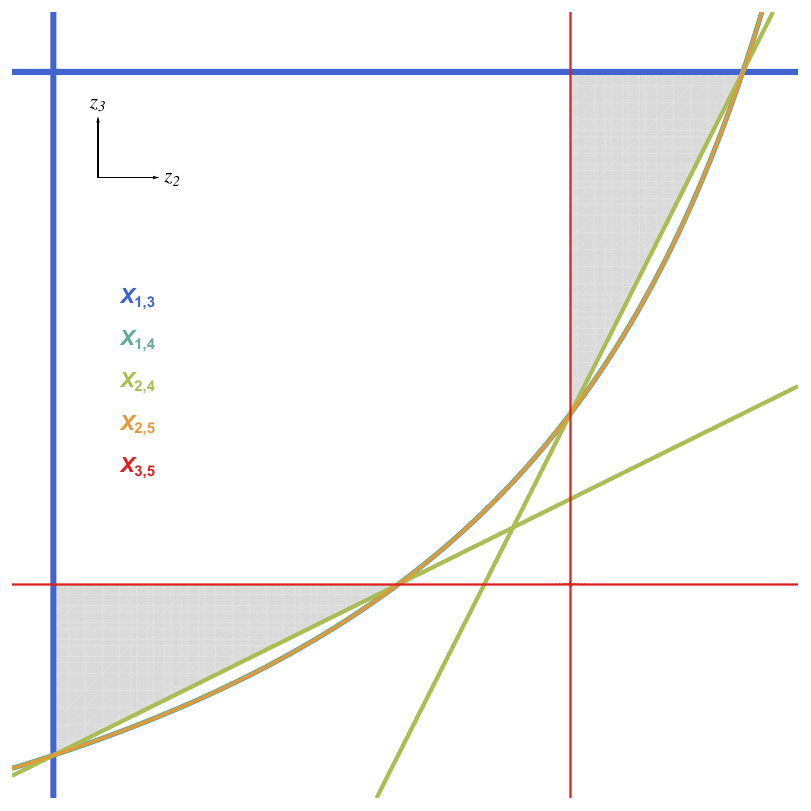}
    \hspace{2cm}
    \includegraphics[width=0.33\textwidth]{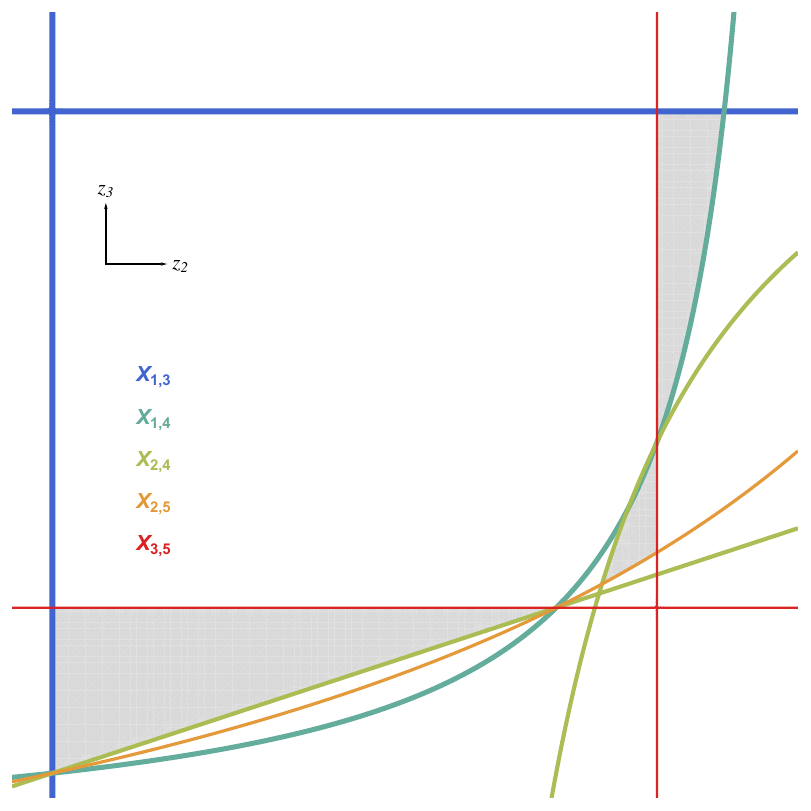}
    \caption{From left to right: Five-point deformed geometry for $(c_{1,3},c_{1,4},c_{2,4})=(2,1,2), (4,2,1)$, respectively. The horizontal axis denotes $z_2$, while the vertical axis denotes $z_3$.}
    \label{fig:deform}
\end{figure*}

The five--point case $n=5$ is slightly more interesting. As in the four--point example, momentum conservation allows us to solve for four variables, for instance, we choose $\omega_1, \omega_2, \omega_3,\omega_4$. In addition, we now impose three ABHY conditions:
\begin{equation}
    \frac{\omega_5^2}{\operatorname{adj} (Q)_{5,5}}2\, Q_{1,3} \operatorname{adj} (Q)_{1,3} = -c_{1,3}, \quad  \frac{\omega_5^2}{\operatorname{adj} (Q)_{5,5}}2\, Q_{1,4} \operatorname{adj} (Q)_{1,4} = -c_{1,4} , \quad  \frac{\omega_5^2}{\operatorname{adj} (Q)_{5,5}}2\, Q_{2,4} \operatorname{adj} (Q)_{2,4} = -c_{2,4} \,,
\end{equation}
from which we can solve for $\omega_5, z_2',z_3'$, leaving a two--dimensional space parametrized by $z_2, z_3$. We then obtain the deformed geometry
\begin{equation}
    \begin{cases}
        &X_{1,3}=\frac{z_2 \left(z_3 c_{1,3}-z_3 c_{1,4}-2 c_{1,3}\right)}{\left(z_2-2\right) z_3} >0 \\
        &X_{1,4}=-\frac{-z_3 c_{1,4}+z_2 z_3 c_{1,4}+z_2 c_{2,4}-z_3 c_{2,4}}{\left(z_2-1\right) \left(z_3-2\right)} >0\\
        &X_{2,4}= \frac{\left(z_2-z_3\right) \left(-4 z_2 c_{1,3}+2 z_2 z_3 c_{1,3}-2 z_3 c_{1,3}-2 z_2 z_3 c_{1,4}+2 z_3 c_{1,4}-z_2 z_3 c_{2,4}+2 z_3 c_{2,4}+4 c_{1,3}\right)}{\left(z_2-2\right) \left(z_2-1\right) \left(z_3-2\right) z_3} >0 \\
        &X_{2,5}= \frac{2 \left(z_2 c_{1,3}-z_3 c_{1,3}+z_2 z_3 c_{1,4}-z_3 c_{1,4}\right)}{\left(z_2-2\right) z_3} >0 \\
        &X_{3,5}=\frac{\left(z_3-1\right) \left(2 z_2 c_{1,4}+z_2 c_{2,4}-2 c_{1,4}-2 c_{2,4}\right)}{\left(z_2-1\right) \left(z_3-2\right)}>0\,,\\
    \end{cases}
\end{equation}
in terms of the five real parameters $z_2,z_3,c_{1,3},c_{1,4},c_{2,4}$. The two-dimensional positive region takes different shapes depending on the values of the $c_{i,j}$'s. Two representative cases are shown in Figure~\ref{fig:deform}.

\noindent

On the left of Figure~\ref{fig:deform}, the positive region consists of two disjoint triangles, each bounded by $X_{1,3}=0$, $X_{2,4}=0$, and $X_{3,5}=0$, while the boundaries associated with $X_{1,4}$ and $X_{2,5}$ collapse to vertices. On the right, the region splits into three components, one triangular and two with curved boundaries, illustrating that the same boundary can be realized on several components. For the first case, the two triangles are specified by
\begin{equation}
\begin{aligned}
    &0<z_2<z_3<1\,,\\
    &\frac{3}{2}<z_2<2\,,\qquad 4z_2-4<z_3<4\,.
\end{aligned}
\end{equation}
With the orientations induced by the map to $(X_{1,3},X_{1,4})$, which are opposite on the two triangles relative to $dz_2\wedge dz_3$, the canonical form of their union reads
\begin{equation}
\begin{aligned}
\Omega_5={}&d\log\left(\frac{z_2}{1-z_2}\right)
    \wedge d\log\left(\frac{z_3-z_2}{1-z_3}\right)-d\log\left(\frac{z_2-\frac{3}{2}}{2-z_2}\right)
    \wedge d\log\left(\frac{z_3-4z_2+4}{4-z_3}\right)\\
={}&\left[\frac{1}{z_2(z_3-z_2)(1-z_3)}
    -\frac{4}{(2z_2-3)(z_3-4z_2+4)(4-z_3)}\right]
    dz_2\wedge dz_3\,.
\end{aligned}
\end{equation}
With the same gauge choice and values of $c_{i,j}$, this expression coincides with $\hat{\Omega}_{5}^{\phi^3}$ as expected.

\section{Scalar-Scaffolded Gluons and Gravitons}\label{sec:scalar-scaffolded}
As mentioned earlier, the method we used to evaluate the Mellin integrals for $\operatorname{tr}(\phi^3)$ amplitudes can be applied just as well to tree-level amplitudes of different theories. In particular, the scalar amplitudes of Yang-Mills-scalar (YMS) and Einstein-Maxwell-scalar (EMS) theories are especially interesting~\cite{Cachazo:2014xea, Cachazo:2015ksa}, since their special components in which the scalars are paired as $(1,2), (3,4),\ldots ,(2n-1,2n)$ yield gluon and graviton amplitudes via the $n$-fold scaffolding residues at $s_{1,2}=s_{3,4}=\ldots =s_{2n-1,2n}=0$ as studied in~\cite{Arkani-Hamed:2023swr,Arkani-Hamed:2023jry,Arkani-Hamed:2024tzl}. Cf. also~\cite{Arkani-Hamed:2008owk,Stieberger:2014cea, Stieberger:2015qja,Cachazo:2015ksa} for similar ideas. 

By applying the same trick, it is straightforward to obtain the integrated YMS/EMS amplitudes:
\begin{equation} 
\begin{aligned}
    &\tilde{\cal A}^{\rm YMS/EMS}_{(1,2)\ldots(2n-1,2n)}(\Delta_i,\mathbf{z}_i) \\
    =& \int_0^\infty d\omega_{2n}\,\frac{ \omega_{2n}^{\sum_i \Delta_i-2n+2m}}{\omega_{2n}^{D-2n+1} \mathcal{J}(\mathbf{z}_i)} \,
    \delta^{(D-2n+1)}\!\left(R_I(\mathbf{z}_i)\right)
    \prod_{a=1}^{2n-1} w_a^{\Delta_a-1}(\mathbf{z}_i)   \hat{{\cal A}}'^{\rm YMS/EMS}_{(1,2)\ldots(2n-1,2n)} (\mathbf{z}_i)\,,
\end{aligned}
\end{equation}
where $m$ is related to the mass-dimension of the $2n$-point amplitudes, which takes the values $m=2-n$ for YMS and $m=1$ for EMS; the prime in $ \hat{{\cal A}}'^{\rm YMS/EMS}_{(1,2)\ldots(2n-1,2n)} (\mathbf{z}_i)$ indicates that we have not stripped off the overall factor $\Lambda^{-m}(\mathbf{z}_i)$, which we keep for later convenience. The factor $\delta^{(D-2n+1)}(R_I(\mathbf{z}_i))$ is the residual momentum-conservation support for the $2n$-point scaffolded amplitude, written in the same patchwise sense as in the main text.
The gluon and graviton amplitudes are then given by:
\begin{equation}
\begin{aligned}
  &\tilde{\cal A}^{\rm YM/GR}_{2n\to n}(\Delta_i,\mathbf{z}_i)\\
  = &  
   g(\omega_i) \Lambda^n(\mathbf{z}_i){\rm Res}_{\hat{s}_{1,2}=...=0}   \tilde{\cal A}^{\rm YMS/EMS}_{(1,2)\ldots(2n-1,2n)}(\Delta_i,\mathbf{z}_i)\,,
\end{aligned}
\end{equation}
where we take residues of the dimensionless variables $\hat{s}_{1,2}/\Lambda(\mathbf{z}_i)=\ldots=\hat{s}_{2n-1,2n}/\Lambda(\mathbf{z}_i)=0$ and multiply by the factor $g(\omega)= \prod_{b \in {\rm odd}} \omega_b$ for YM and $g(\omega)= \prod_{b \in {\rm odd}} \omega_b^2$ for GR, prior to performing the integrals, to precisely match the usual YM/GR amplitudes, as explained below. We also define:
\begin{equation}
    \hat{\cal A}'^{\rm YM/GR}_{2n\to n}(\mathbf{z}_i):=\Lambda^n(\mathbf{z}_i) {\rm Res}_{\hat{s}_{1,2}=...=0}  \hat{\cal A}'^{\rm YMS/EMS}_{(1,2)\ldots(2n-1,2n)}(\mathbf{z}_i)\,.
\end{equation}

The above construction is equivalent to the Mellin transform on the YM/GR amplitudes by performing the integral using solution~\eqref{eq: solution omega}:
\begin{equation} 
\begin{aligned}
    &\tilde{\cal A}^{\rm YM/GR}_n(\Delta_i,\mathbf{z}_i) \\
    =& \int_0^\infty d\omega_n\, \frac{ \omega_n^{\sum_i \Delta_i-n+m'}}{\omega_n^{D-n+1} \mathcal{J}(\mathbf{z}_i)} \,
    \delta^{(D-n+1)}\!\left(R_I(\mathbf{z}_i)\right)
    \prod_{a=1}^{n-1} w_a^{\Delta_a-1}(\mathbf{z}_i)   \hat{{\cal A}}'^{\rm YM/GR}_n (\mathbf{z}_i) \\
\end{aligned}
\end{equation}
with $m'=4-n$ for YM and $m'=2$ for GR; here $\hat{{\cal A}}'^{\rm YM/GR}_n (\mathbf{z}_i)$ denotes the YM/GR amplitudes (without stripping off $\Lambda(\mathbf{z}_i)$), evaluated under the following replacements (to distinguish from the $2n$-scalar momenta we write $k_i^\mu=\omega_i v_i^\mu$ for gluon momenta):
\begin{equation*}
    \varepsilon_a\cdot k_b \to  \frac{\operatorname{adj}(Q)_{b,n}}{\Lambda (\mathbf{z}_i)}  \varepsilon_a \cdot v_b
\end{equation*}
\vspace{-0.7cm}
\begin{equation*}
    k_a\cdot k_b \to \frac{ \operatorname{adj}(Q)_{a,b}}{\Lambda(\mathbf{z}_i)}  v_a \cdot v_b \quad\text{or}\quad  \frac{ \operatorname{adj}(Q)_{a,n}\operatorname{adj}(Q)_{b,n}}{\Lambda^2(\mathbf{z}_i)}  v_a \cdot v_b \,.
\end{equation*} 
We also provide the simplest example of 3-gluon and $6\to3$ scalar-scaffolded amplitudes in the following subsection.


\subsection{The exact correspondence between the scalar-scaffolded and the standard YM/GR}
In this subsection, we provide additional details on the scalar-scaffolded Yang-Mills amplitude and its correspondence with the usual YM amplitude (the EMS-scaffolded GR case follows exactly the same argument). As it was studied in~\cite{Arkani-Hamed:2023swr,Arkani-Hamed:2023jry,Arkani-Hamed:2024tzl}, for amplitudes in the momentum space we have:
\begin{equation} \label{eq: momentum scaffolding}
\begin{aligned}
  {\cal A}^{\rm YM}_{2n\to n}
  = {\rm Res}_{s_{1,2}=s_{3,4}=...=s_{2n-1,2n}=0}  {\cal A}^{\rm YMS}_{(1,2)\ldots(2n-1,2n)}\,,
\end{aligned}
\end{equation}
where the momentum of the $i$-th gluon $k_i^\mu$ is given by 
\begin{equation}\label{eq:momenta to 2n-gon}
    k_i^\mu=(p_{2i-1}+p_{2i})^\mu.
\end{equation}
The polarization, on the other hand, contains a free parameter $\alpha$, which is associated with the gauge redundancy.
\begin{equation}\label{eq:polarization to 2n-gon}
    \varepsilon_i^\mu=(1-\alpha)p_{2i-1}^\mu-\alpha p_{2i}^\mu,
\end{equation}
which one can verify it satisfying the transverality condition $\varepsilon_i\cdot k_i=0$ and $\varepsilon_i\cdot\varepsilon_i=0$. We note that the gluon on-shell condition
\begin{equation}
    k_{i}^2= 2 \omega_{2i-1} \omega_{2i}\, q_{2i-1}\cdot q_{2i} =0
\end{equation}
which holds on the scaffolding residues, should be realized through the collinear limit $q_{2i-1}\cdot q_{2i}=0$ rather than through a soft limit $\omega_{2i-1}=0$ or $\omega_{2i}=0$. Otherwise, one encounters unwanted conditions such as $s_{1,3}=0$. Under this $2n$-point collinear limit, the kinematic system degenerates to an effective $n$-point configuration. Moreover, the precise map between the scaffolded planar (scalar) variables and the usual inner products of $p,\varepsilon$ is given by
\begin{align}  \label{eq: eptrans}
&2 \varepsilon_i \cdot \varepsilon_j=X_{2 i,2 j-1}+ X_{2 i-1,2 j}- X_{2 i-1,2 j-1}-X_{2 i,2 j}\,, \nonumber  \\ 
&2 \varepsilon_i \cdot k_j=X_{2i, 2j{+}1}-X_{2i, 2j{-}1}+X_{2i{-}1, 2j{-}1}-X_{2i{-}1, 2j{+}1} \,, \\ 
 &2 k_i \cdot k_j= X_{2 i-1,2 j+1}+X_{2 i+1,2 j-1}- X_{2 i-1,2 j-1} - X_{2 i+1,2 j+1}\,, \nonumber
\end{align}
where we have chosen $\alpha=1$. The above translation is sloppy at the level of mass dimensions for our purpose. In fact, comparing the mass dimensions (defined as the power of momentum) of the scaffolded amplitudes ${\cal A}^{\rm YM}_{2n\to n}$ and the usual ${\cal A}^{\rm YM}_{n}$, we find 
\begin{equation}
    [{\cal A}^{\rm YM}_{2n\to n}]=4 \quad \text{and} \quad [{\cal A}^{\rm YM}_{n}]=4-n \,.
\end{equation}
This discrepancy arises because in going from the left-hand side to the right-hand side of Eq.~\eqref{eq: eptrans}, the the mass-dimension of each polarization vector $\varepsilon$ is increased by one.

Now we introduce the following dimensionless transformation (with $k_i^\mu=\omega_i v_i^\mu$ as our definition):
\begin{align}  \label{eq: eptrans2}
& \Lambda^{(2n)}(\mathbf{z}_i)\,2 \varepsilon_i \cdot \varepsilon_j=\hat{X}_{2 i,2 j-1}+ \hat{X}_{2 i-1,2 j}- \hat{X}_{2 i-1,2 j-1}-\hat{X}_{2 i,2 j}\,, \nonumber  \\ 
& \frac{\Lambda^{(2n)}(\mathbf{z}_i)}{\Lambda^{(n)}(\mathbf{z}_i)} \operatorname{adj}(Q)_{j,n}\,2 \varepsilon_i \cdot v_j =\hat{X}_{2i, 2j{+}1}-\hat{X}_{2i, 2j{-}1}+\hat{X}_{2i{-}1, 2j{-}1}-\hat{X}_{2i{-}1, 2j{+}1} \,, \\ 
 & \frac{\Lambda^{(2n)}(\mathbf{z}_i)}{\Lambda^{(n)}(\mathbf{z}_i)}\operatorname{adj}(Q)_{i,j}\,2 v_i \cdot v_j= \hat{X}_{2 i-1,2 j+1}+\hat{X}_{2 i+1,2 j-1}- \hat{X}_{2 i-1,2 j-1} - \hat{X}_{2 i+1,2 j+1}\,, \nonumber
\end{align}
where under the identification $\omega_{2i-1}^{(2n)}, \omega_{2i}^{(2n)}\to \omega_{i}^{(n)}$ the LHS differ from those in~\eqref{eq: eptrans} by an overall factor of $\omega^{(n)}_n$ for each polarization vector. Using the above transformation, we obtain
\begin{equation}
    \hat{\cal A}'^{\rm YM}_{2n\to n}(\mathbf{z}_i):=\Lambda^n(\mathbf{z}_i)\, {\rm Res}_{\hat{s}_{1,2}=...=0}  \hat{\cal A}'^{\rm YMS}_{(1,2)\ldots(2n-1,2n)}(\mathbf{z}_i) =  \hat{\cal A}'^{\rm YM}_{n}(\mathbf{z}_i)\,.
\end{equation}
Meanwhile, we have
\begin{equation}
    {\cal A}^{\rm YM}_{2n\to n}=\omega_{2n}^{4-2n} \hat{\cal A}'^{\rm YM}_{2n\to n}
\end{equation}
where we have the factor $\omega_{2n}^{4-2n}$ (instead of $\omega_{2n}^{4}$) appears because the RHS is evaluated on dimensionless residues. Likewise,
\begin{equation}
    {\cal A}^{\rm YM}_{n}=\omega_{n}^{4-n} \hat{\cal A}'^{\rm YM}_{n} \,.
\end{equation}
Therefore, under the dimensionless transformation we obtain
\begin{equation}
   \omega_{2n}^{n} {\cal A}^{\rm YM}_{2n\to n}\xleftrightarrow[]{\eqref{eq: eptrans2}} {\cal A}^{\rm YM}_{ n}\,,
\end{equation}
where the additional factor of $\omega$ should be included before performing the integrals. However, since the Mellin transform involves the factors $\prod_a \omega_a^{\Delta_a-1}$, and its mass dimension is further modified by the Jacobian, it is more convenient to use
\begin{equation}
   \prod_{b\in {\rm odd}}\omega_b\, \tilde{\cal A}^{\rm YM}_{2n\to n}\xleftrightarrow[]{\eqref{eq: eptrans2}} \tilde{\cal A}^{\rm YM}_{ n}\,.
\end{equation}
This correspondence holds under the identifications $\omega_{2i-1}^{(2n)}, \omega_{2i}^{(2n)}\to \omega_{i}^{(n)}$ and $ \Delta_{2i-1}^{(2n)},\Delta_{2i}^{(2n)} \to 1/2\Delta_{i}^{(n)}$.

Let us consider the simplest $3$-gluon amplitude, for the $6 \to 3$ scalar-scaffolded amplitude we have:
\begin{equation} \label{eq: 6to3pt YM}
\begin{aligned}
  \tilde{\cal A}^{\rm YM}_{6\to 3}(\Delta_i,\mathbf{z}_i)
  = 
   \int_0^\infty d\omega_{6}\,\frac{ \omega_{6}^{\sum_i \Delta_i-D-3}}{ \mathcal{J}(\mathbf{z}_i)}
   \delta^{(D-5)}\!\left(R_I(\mathbf{z}_i)\right)
   \omega_6^3 w_1 w_3 w_5 \, \prod_{a=1}^{5} w_a^{\Delta_a-1}(\mathbf{z}_i)    \hat{\cal A}'^{\rm YM}_{6\to 3} (\mathbf{z}_i)\,,
\end{aligned}
\end{equation}
with
\begin{equation*}
\begin{aligned}
    \hat{{\cal A}}'^{\rm YM}_{6\to 3} (\mathbf{z}_i)=-\frac{1}{\Lambda^{2}(\mathbf{z}_i)}\left((\hat{s}_{2,3,4}-\hat{s}_{1,6})\hat{s}_{1,2,3} \right.\left. +(\hat{s}_{3,4,5}-\hat{s}_{4,5})\hat{s}_{2,3.4}+(\hat{s}_{1,2,3}-\hat{s}_{2,3})\hat{s}_{3,4,5}\right)
\end{aligned}
\end{equation*}
And similarly for the (non-scaffolded) 3-point amplitude we have
\begin{equation} \label{eq: 3pt YM}
\begin{aligned}
    \tilde{\cal A}^{\rm YM}_3(\Delta_i,\mathbf{z}_i) 
    = \int_0^\infty d\omega_3\, \frac{ \omega_3^{\sum_i \Delta_i-D}}{ \mathcal{J}(\mathbf{z}_i)} \,
    \delta^{(D-2)}\!\left(R_I(\mathbf{z}_i)\right)
    \prod_{a=1}^{2} w_a^{\Delta_a-1}(\mathbf{z}_i)   \hat{{\cal A}}'^{\rm YM/GR}_3 (\mathbf{z}_i) \,,
\end{aligned}
\end{equation}
with 
\begin{equation*}
\begin{aligned}
\hat{{\cal A}}'^{\rm YM}_3 (\mathbf{z}_i)= \frac{-4}{\Lambda (\mathbf{z}_i)}\left(\operatorname{adj}(Q)_{3,3}\, \varepsilon_1 \cdot \varepsilon_3\; \varepsilon_2 \cdot v_3 \right.  \left. +\operatorname{adj}(Q)_{2,3}\, \varepsilon_2 \cdot \varepsilon_3\; \varepsilon_1 \cdot v_2+\operatorname{adj}(Q)_{1,3}\, \varepsilon_1 \cdot \varepsilon_2\; \varepsilon_3\cdot v_1\right)
\end{aligned}
\end{equation*}
Quite nicely, using the translation we describe above (see also~\cite{Arkani-Hamed:2023jry}) and the identification $\omega_{2i-1}^{(2n)}, \omega_{2i}^{(2n)}\to \omega_{i}^{(n)}$ and $ \Delta_{2i-1}^{(2n)},\Delta_{2i}^{(2n)} \to 1/2\Delta_{i}^{(n)}$ in~\eqref{eq: 6to3pt YM}, the expression matches~\eqref{eq: 3pt YM}~\footnote{One should be careful with the Jacobian factor in scalar-scaffolded amplitudes. Under scaffolded collinear kinematics, the Jacobian vanishes; therefore, one must approach the collinear limit with care and factor out the resulting divergence.}. 

~\\
\twocolumngrid

\bibliographystyle{apsrev4-2.bst}
\bibliography{Refs.bib}

@article{Donnay:2021wrk,
    author = "Donnay, Laura and Ruzziconi, Romain",
    title = "{BMS flux algebra in celestial holography}",
    eprint = "2108.11969",
    archivePrefix = "arXiv",
    primaryClass = "hep-th",
    doi = "10.1007/JHEP11(2021)040",
    journal = "JHEP",
    volume = "11",
    pages = "040",
    year = "2021"
}

@article{Mason:2023mti,
    author = "Mason, Lionel and Ruzziconi, Romain and Yelleshpur Srikant, Akshay",
    title = "{Carrollian amplitudes and celestial symmetries}",
    eprint = "2312.10138",
    archivePrefix = "arXiv",
    primaryClass = "hep-th",
    doi = "10.1007/JHEP05(2024)012",
    journal = "JHEP",
    volume = "05",
    pages = "012",
    year = "2024"
}

@article{Hu:2021lrx,
    author = "Hu, Yangrui and Ren, Lecheng and Srikant, Akshay Yelleshpur and Volovich, Anastasia",
    title = "{Celestial dual superconformal symmetry, MHV amplitudes and differential equations}",
    eprint = "2106.16111",
    archivePrefix = "arXiv",
    primaryClass = "hep-th",
    doi = "10.1007/JHEP12(2021)171",
    journal = "JHEP",
    volume = "12",
    pages = "171",
    year = "2021"
}

@incollection{ZagierMellin,
  author       = {Zagier, Don},
  title        = {The Mellin Transform and Other Useful Analytic Techniques},
  booktitle    = {Quantum Field Theory and Modular Forms},
  series       = {Quanta of Maths},
  publisher    = {American Mathematical Society},
  year         = {2009},
  pages        = {1--23},
  editor       = {Ramakrishnan, Dinakar and 
                  Murty, M. Ram and 
                  Saha, Kumares},
  doi          = {10.1090/pspum/059},
  note         = {Originally appeared as: 
                  {\em Appendix to E. Zeidler, 
                  Quantum Field Theory I: Basics in Mathematics and Physics}, 
                  Springer (2006)},
}

@article{Costello:2022jpg,
    author = "Costello, Kevin and Paquette, Natalie M. and Sharma, Atul",
    title = "{Top-Down Holography in an Asymptotically Flat Spacetime}",
    eprint = "2208.14233",
    archivePrefix = "arXiv",
    primaryClass = "hep-th",
    doi = "10.1103/PhysRevLett.130.061602",
    journal = "Phys. Rev. Lett.",
    volume = "130",
    number = "6",
    pages = "061602",
    year = "2023"
}

@article{deBoer:2003vf,
    author = "de Boer, Jan and Solodukhin, Sergey N.",
    title = "{A Holographic reduction of Minkowski space-time}",
    eprint = "hep-th/0303006",
    archivePrefix = "arXiv",
    reportNumber = "ITFA-2003-11",
    doi = "10.1016/S0550-3213(03)00494-2",
    journal = "Nucl. Phys. B",
    volume = "665",
    pages = "545--593",
    year = "2003"
}

@book{Strominger:2017zoo,
    author = "Strominger, Andrew",
    title = "{Lectures on the Infrared Structure of Gravity and Gauge Theory}",
    eprint = "1703.05448",
    archivePrefix = "arXiv",
    primaryClass = "hep-th",
    isbn = "978-0-691-17973-5",
    publisher = "Princeton University Press",
    year = "2018"
}

@article{Pasterski:2016qvg,
    author = "Pasterski, Sabrina and Shao, Shu-Heng and Strominger, Andrew",
    title = "{Flat Space Amplitudes and Conformal Symmetry of the Celestial Sphere}",
    eprint = "1701.00049",
    archivePrefix = "arXiv",
    primaryClass = "hep-th",
    doi = "10.1103/PhysRevD.96.065026",
    journal = "Phys. Rev. D",
    volume = "96",
    number = "6",
    pages = "065026",
    year = "2017"
}

@article{Pasterski:2017kqt,
    author = "Pasterski, Sabrina and Shao, Shu-Heng",
    title = "{Conformal basis for flat space amplitudes}",
    eprint = "1705.01027",
    archivePrefix = "arXiv",
    primaryClass = "hep-th",
    doi = "10.1103/PhysRevD.96.065022",
    journal = "Phys. Rev. D",
    volume = "96",
    number = "6",
    pages = "065022",
    year = "2017"
}

@article{Barnich:2010eb,
    author = "Barnich, Glenn and Troessaert, Cedric",
    title = "{Aspects of the BMS/CFT correspondence}",
    eprint = "1001.1541",
    archivePrefix = "arXiv",
    primaryClass = "hep-th",
    reportNumber = "ULB-TH-09-28",
    doi = "10.1007/JHEP05(2010)062",
    journal = "JHEP",
    volume = "05",
    pages = "062",
    year = "2010"
}

@article{Bagchi:2010zz,
    author = "Bagchi, Arjun",
    title = "{Correspondence between Asymptotically Flat Spacetimes and Nonrelativistic Conformal Field Theories}",
    eprint = "1006.3354",
    archivePrefix = "arXiv",
    primaryClass = "hep-th",
    doi = "10.1103/PhysRevLett.105.171601",
    journal = "Phys. Rev. Lett.",
    volume = "105",
    pages = "171601",
    year = "2010"
}

@article{Kervyn:2025wsb,
    author = "Kervyn, Xavier and Stieberger, Stephan",
    title = "{High energy string theory and the celestial sphere}",
    eprint = "2504.13738",
    archivePrefix = "arXiv",
    primaryClass = "hep-th",
    doi = "10.1007/JHEP09(2025)044",
    journal = "JHEP",
    volume = "09",
    pages = "044",
    year = "2025"
}

@article{Schreiber:2017jsr,
    author = "Schreiber, Anders and Volovich, Anastasia and Zlotnikov, Michael",
    title = "{Tree-level gluon amplitudes on the celestial sphere}",
    eprint = "1711.08435",
    archivePrefix = "arXiv",
    primaryClass = "hep-th",
    doi = "10.1016/j.physletb.2018.04.010",
    journal = "Phys. Lett. B",
    volume = "781",
    pages = "349--357",
    year = "2018"
}

@article{Nandan:2019jas,
    author = "Nandan, Dhritiman and Schreiber, Anders and Volovich, Anastasia and Zlotnikov, Michael",
    title = "{Celestial Amplitudes: Conformal Partial Waves and Soft Limits}",
    eprint = "1904.10940",
    archivePrefix = "arXiv",
    primaryClass = "hep-th",
    doi = "10.1007/JHEP10(2019)018",
    journal = "JHEP",
    volume = "10",
    pages = "018",
    year = "2019"
}

@article{Kulkarni:2025qcx,
    author = "Kulkarni, Harshal and Ruzziconi, Romain and Yelleshpur Srikant, Akshay",
    title = "{On Carrollian and celestial correlators in general dimensions}",
    eprint = "2508.06602",
    archivePrefix = "arXiv",
    primaryClass = "hep-th",
    doi = "10.1007/JHEP10(2025)187",
    journal = "JHEP",
    volume = "10",
    pages = "187",
    year = "2025"
}

@article{Stieberger:2018onx,
    author = "Stieberger, Stephan and Taylor, Tomasz R.",
    title = "{Symmetries of Celestial Amplitudes}",
    eprint = "1812.01080",
    archivePrefix = "arXiv",
    primaryClass = "hep-th",
    reportNumber = "MPP-2018-292",
    doi = "10.1016/j.physletb.2019.03.063",
    journal = "Phys. Lett. B",
    volume = "793",
    pages = "141--143",
    year = "2019"
}

@article{Cachazo:2013gna,
    author = "Cachazo, Freddy and He, Song and Yuan, Ellis Ye",
    title = "{Scattering equations and Kawai-Lewellen-Tye orthogonality}",
    eprint = "1306.6575",
    archivePrefix = "arXiv",
    primaryClass = "hep-th",
    doi = "10.1103/PhysRevD.90.065001",
    journal = "Phys. Rev. D",
    volume = "90",
    number = "6",
    pages = "065001",
    year = "2014"
}

@article{Stieberger:2018edy,
    author = "Stieberger, Stephan and Taylor, Tomasz R.",
    title = "{Strings on Celestial Sphere}",
    eprint = "1806.05688",
    archivePrefix = "arXiv",
    primaryClass = "hep-th",
    reportNumber = "MPP-2018-136",
    doi = "10.1016/j.nuclphysb.2018.08.019",
    journal = "Nucl. Phys. B",
    volume = "935",
    pages = "388--411",
    year = "2018"
}

@article{Stieberger:2015qja,
    author = "Stieberger, Stephan and Taylor, Tomasz R.",
    title = "{Graviton Amplitudes from Collinear Limits of Gauge Amplitudes}",
    eprint = "1502.00655",
    archivePrefix = "arXiv",
    primaryClass = "hep-th",
    reportNumber = "MPP-2015-001",
    doi = "10.1016/j.physletb.2015.03.053",
    journal = "Phys. Lett. B",
    volume = "744",
    pages = "160--162",
    year = "2015"
}

@article{Donnay:2022aba,
    author = "Donnay, Laura and Fiorucci, Adrien and Herfray, Yannick and Ruzziconi, Romain",
    title = "{Carrollian Perspective on Celestial Holography}",
    eprint = "2202.04702",
    archivePrefix = "arXiv",
    primaryClass = "hep-th",
    doi = "10.1103/PhysRevLett.129.071602",
    journal = "Phys. Rev. Lett.",
    volume = "129",
    number = "7",
    pages = "071602",
    year = "2022"
}

@article{Donnay:2022wvx,
    author = "Donnay, Laura and Fiorucci, Adrien and Herfray, Yannick and Ruzziconi, Romain",
    title = "{Bridging Carrollian and celestial holography}",
    eprint = "2212.12553",
    archivePrefix = "arXiv",
    primaryClass = "hep-th",
    doi = "10.1103/PhysRevD.107.126027",
    journal = "Phys. Rev. D",
    volume = "107",
    number = "12",
    pages = "126027",
    year = "2023"
}

@article{Castiblanco:2024hnq,
    author = "Castiblanco, Lina and Giribet, Gaston and Marin, Gabriel and Rojas, Francisco",
    title = "{Celestial strings: Field theory, conformally soft limits, and mapping the~worldsheet onto the celestial sphere}",
    eprint = "2405.01643",
    archivePrefix = "arXiv",
    primaryClass = "hep-th",
    doi = "10.1103/PhysRevD.110.126001",
    journal = "Phys. Rev. D",
    volume = "110",
    number = "12",
    pages = "126001",
    year = "2024"
}

@article{Arkani-Hamed:2017mur,
    author = "Arkani-Hamed, Nima and Bai, Yuntao and He, Song and Yan, Gongwang",
    title = "{Scattering Forms and the Positive Geometry of Kinematics, Color and the Worldsheet}",
    eprint = "1711.09102",
    archivePrefix = "arXiv",
    primaryClass = "hep-th",
    doi = "10.1007/JHEP05(2018)096",
    journal = "JHEP",
    volume = "05",
    pages = "096",
    year = "2018"
}

@article{Cachazo:2013hca,
    author = "Cachazo, Freddy and He, Song and Yuan, Ellis Ye",
    title = "{Scattering of Massless Particles in Arbitrary Dimensions}",
    eprint = "1307.2199",
    archivePrefix = "arXiv",
    primaryClass = "hep-th",
    doi = "10.1103/PhysRevLett.113.171601",
    journal = "Phys. Rev. Lett.",
    volume = "113",
    number = "17",
    pages = "171601",
    year = "2014"
}

@article{Cachazo:2013iea,
    author = "Cachazo, Freddy and He, Song and Yuan, Ellis Ye",
    title = "{Scattering of Massless Particles: Scalars, Gluons and Gravitons}",
    eprint = "1309.0885",
    archivePrefix = "arXiv",
    primaryClass = "hep-th",
    doi = "10.1007/JHEP07(2014)033",
    journal = "JHEP",
    volume = "07",
    pages = "033",
    year = "2014"
}

@article{Cachazo:2014xea,
    author = "Cachazo, Freddy and He, Song and Yuan, Ellis Ye",
    title = "{Scattering Equations and Matrices: From Einstein To Yang-Mills, DBI and NLSM}",
    eprint = "1412.3479",
    archivePrefix = "arXiv",
    primaryClass = "hep-th",
    doi = "10.1007/JHEP07(2015)149",
    journal = "JHEP",
    volume = "07",
    pages = "149",
    year = "2015"
}

@article{Arkani-Hamed:2024tzl,
    author = "Arkani-Hamed, Nima and Cao, Qu and Dong, Jin and Figueiredo, Carolina and He, Song",
    title = "{Surface Kinematics and the Canonical Yang-Mills All-Loop Integrand}",
    eprint = "2408.11891",
    archivePrefix = "arXiv",
    primaryClass = "hep-th",
    doi = "10.1103/PhysRevLett.134.171601",
    journal = "Phys. Rev. Lett.",
    volume = "134",
    number = "17",
    pages = "171601",
    year = "2025"
}

@article{Arkani-Hamed:2023swr,
    author = "Arkani-Hamed, Nima and Cao, Qu and Dong, Jin and Figueiredo, Carolina and He, Song",
    title = "{Hidden zeros for particle/string amplitudes and the unity of colored scalars, pions and gluons}",
    eprint = "2312.16282",
    archivePrefix = "arXiv",
    primaryClass = "hep-th",
    doi = "10.1007/JHEP10(2024)231",
    journal = "JHEP",
    volume = "10",
    pages = "231",
    year = "2024"
}

@article{Arkani-Hamed:2023jry,
    author = "Arkani-Hamed, Nima and Cao, Qu and Dong, Jin and Figueiredo, Carolina and He, Song",
    title = "{Scalar-scaffolded gluons and the combinatorial origins of Yang-Mills theory}",
    eprint = "2401.00041",
    archivePrefix = "arXiv",
    primaryClass = "hep-th",
    doi = "10.1007/JHEP04(2025)078",
    journal = "JHEP",
    volume = "04",
    pages = "078",
    year = "2025"
}

@article{Stieberger:2014cea,
    author = "Stieberger, Stephan and Taylor, Tomasz R.",
    title = "{Graviton as a Pair of Collinear Gauge Bosons}",
    eprint = "1409.4771",
    archivePrefix = "arXiv",
    primaryClass = "hep-th",
    reportNumber = "MPP-2014-343",
    doi = "10.1016/j.physletb.2014.10.057",
    journal = "Phys. Lett. B",
    volume = "739",
    pages = "457--461",
    year = "2014"
}

@article{Cachazo:2015ksa,
    author = "Cachazo, Freddy and He, Song and Yuan, Ellis Ye",
    title = "{New Double Soft Emission Theorems}",
    eprint = "1503.04816",
    archivePrefix = "arXiv",
    primaryClass = "hep-th",
    doi = "10.1103/PhysRevD.92.065030",
    journal = "Phys. Rev. D",
    volume = "92",
    number = "6",
    pages = "065030",
    year = "2015"
}

@article{Arkani-Hamed:2008owk,
    author = "Arkani-Hamed, Nima and Cachazo, Freddy and Kaplan, Jared",
    title = "{What is the Simplest Quantum Field Theory?}",
    eprint = "0808.1446",
    archivePrefix = "arXiv",
    primaryClass = "hep-th",
    doi = "10.1007/JHEP09(2010)016",
    journal = "JHEP",
    volume = "09",
    pages = "016",
    year = "2010"
}

@article{Mizera:2022sln,
    author = "Mizera, Sebastian and Pasterski, Sabrina",
    title = "{Celestial geometry}",
    eprint = "2204.02505",
    archivePrefix = "arXiv",
    primaryClass = "hep-th",
    doi = "10.1007/JHEP09(2022)045",
    journal = "JHEP",
    volume = "09",
    pages = "045",
    year = "2022"
}

@article{Melton:2023bjw,
    author = "Melton, Walker and Sharma, Atul and Strominger, Andrew",
    title = "{Celestial leaf amplitudes}",
    eprint = "2312.07820",
    archivePrefix = "arXiv",
    primaryClass = "hep-th",
    doi = "10.1007/JHEP07(2024)132",
    journal = "JHEP",
    volume = "07",
    pages = "132",
    year = "2024"
}

@article{Arkani-Hamed:2017tmz,
    author = "Arkani-Hamed, Nima and Bai, Yuntao and Lam, Thomas",
    title = "{Positive Geometries and Canonical Forms}",
    eprint = "1703.04541",
    archivePrefix = "arXiv",
    primaryClass = "hep-th",
    doi = "10.1007/JHEP11(2017)039",
    journal = "JHEP",
    volume = "11",
    pages = "039",
    year = "2017"
}

@article{Casali:2022fro,
    author = "Casali, Eduardo and Melton, Walker and Strominger, Andrew",
    title = "{Celestial amplitudes as AdS-Witten diagrams}",
    eprint = "2204.10249",
    archivePrefix = "arXiv",
    primaryClass = "hep-th",
    doi = "10.1007/JHEP11(2022)140",
    journal = "JHEP",
    volume = "11",
    pages = "140",
    year = "2022"
}

@article{Atanasov:2021oyu,
    author = "Atanasov, Alexander and Ball, Adam and Melton, Walker and Raclariu, Ana-Maria and Strominger, Andrew",
    title = "{(2, 2) Scattering and the celestial torus}",
    eprint = "2101.09591",
    archivePrefix = "arXiv",
    primaryClass = "hep-th",
    doi = "10.1007/JHEP07(2021)083",
    journal = "JHEP",
    volume = "07",
    pages = "083",
    year = "2021"
}

@article{Melton:2024jyq,
    author = "Melton, Walker and Sharma, Atul and Strominger, Andrew",
    title = "{Soft algebras for leaf amplitudes}",
    eprint = "2402.04150",
    archivePrefix = "arXiv",
    primaryClass = "hep-th",
    doi = "10.1007/JHEP07(2024)070",
    journal = "JHEP",
    volume = "07",
    pages = "070",
    year = "2024"
}

\end{document}